\documentclass[12pt,twocolumn,preprint,a4]{iopart}
\usepackage{graphicx}
\usepackage{booktabs}
\usepackage{textcomp}
\usepackage{color}
\usepackage{multirow}
\usepackage{multicol}
\usepackage{enumerate}

\begin{document}

\title[Origin of the energetic ions at the substrate during HiPIMS of titanium]
	{Origin of the energetic ions at the substrate generated during high power pulsed magnetron sputtering of titanium}

\author{C. Maszl\footnote{\ead{christian.maszl@rub.de}}, W. Breilmann, J. Benedikt and A. von Keudell}

\address{Research Department Plasmas with Complex Interactions,
Ruhr-Universit\"at Bochum, Institute for Experimental Physics II,
D-44780 Bochum, Germany}

\date{\today}

\begin{abstract}
High power impulse magnetron sputtering (HiPIMS) plasmas generate energetic metal ions at the substrate as a major difference to conventional direct current magnetron sputtering (dcMS). The origin of these very energetic ions in HiPIMS is still an open issue, which is unraveled by using two fast diagnostics: time resolved mass spectrometry with a temporal resolution of 2~$\mu$s and phase resolved optical emission spectroscopy with a temporal resolution of 1~$\mu$s. A power scan from dcMS-like to HiPIMS plasmas was performed, with a 2-inch magnetron and a titanium target as sputter source and argon as working gas. Clear differences in the transport as well in the energetic properties of Ar$^{+}$, Ar$^{2+}$, Ti$^+$ and Ti$^{2+}$ were observed. For discharges with highest peak power densities a high energetic group of Ti$^+$ and Ti$^{2+}$ could be identified with energies of approximately 25~eV and of 50~eV, respectively. A cold group of ions is always present. It is found that hot ions are observed only, when the plasma enters the spokes regime, which can be monitored by oscillations in the IV-characteristics in the MHz range that are picked up by the used VI-probes. These oscillations are correlated with the spokes phenomenon and are explained as an amplification of the Hall current inside the spokes as hot ionization zones. To explain the presence of energetic ions, we propose a double layer (DL) confining the hot plasma inside a spoke: if an atom becomes ionized inside the spokes region it is accelerated because of the DL to higher energies  whereas its energy remains unchanged if it is ionized outside. In applying this DL model to our measurements the observed phenomena as well as several measurements from other groups can be explained. Only if spokes and a double layer are present the confined particles can gain enough energy to leave the magnetic trap.  We conclude from our findings that the spoke phenomenon represents the essence of HiPIMS plasmas, explaining their good performance for material synthesis applications. 
\end{abstract}

\maketitle

\section{Introduction}

High power impulse magnetron sputtering (HiPIMS) plasmas are magnetron discharges operated in a pulsed mode. During a typical pulse of few tens of $\mu$s, peak power densities of several kW/cm$^{-2}$ are reached on the target producing very high plasma densities, a high ionization degree and an energetic ion flux at the substrate. The contribution of energetic ions to the growth flux improves thin film properties and distinguishes HiPIMS plasmas from conventional Direct Current Magnetron Sputtering (dcMS). Duty cycles of only a few percent or less are permissible to limit the thermal load on the target to ensure its structural integrity. HiPIMS is nowadays a promising technique for numerous applications \cite{Sarakinos2010}.

The employed pulsing schemes in HiPIMS have the disadvantage that the growth rate is smaller compared to dcMS at same average powers. The origin for the low deposition rate is hidden in the very complex dynamic of an intense HiPIMS plasma. During a short pulse, the characteristics of the plasma continuously evolves without necessarily reaching a steady state situation. This aggravates the experiments and demands for fast diagnostics with excellent temporal resolution.  

A HiPIMS pulse can be divided into five phases \cite{Gudmundsson2012}: (i) in the beginning at $t_0=0$~s a negative voltage is applied to the target in a noble gas like argon as plasma forming gas. During the very first microseconds of the pulse electrical breakdown occurs and the plasma potential in front of the target becomes very negative. Potential drops of the order of $-200$~V \cite{Mishra2010,Mishra2011} are reported at the very beginning of the pulse. A significant fraction of ions is trapped in the region of closed magnetic field lines above the target. These unmagnetized ions are bound to the magnetized electrons in the magnetic trap via Coulomb forces. The ion flux to the substrate is low. Only ions from the tail of the Ion Energy Distribution Function (IEDF) or ions which are created outside the magnetic trap can reach the substrate. The gas kinetic temperature increases up to 1200~K \cite{Lundin2011,Lundin2009}; (ii) the second phase is dominated by a strong increase in current. Above the target a torus shaped plasma is developing generating a typical racetrack by target erosion. Almost the whole voltage drop of the discharge falls off in the sheath in front of the target; (iii) in the third phase the composition of the plasma starts to change. Momentum transfer from the sputtered species to the working gas causes a decrease of the neutral gas density. During this rarefaction process the metallicity of the plasma in the magnetic trap increases \cite{Hala2010}; (iv) the evolution of the plasma in phase four can reach a steady state or enter the runaway regime due to self sputtering depending on an intimate balance between the ionization probability, the probability of ions to return to the target and the self-sputtering yield \cite{Anders2011}; (v) The fifth phase is the afterglow phase. Energetic electrons are typically lost within the first 30~$\mu$s which also causes a strong lowering of the electron temperature. The decay of the cold group of electrons can take ms \cite{Poolcharuansin2010}. 

Transport properties of HiPIMS plasmas have been investigated by numerous contributors by monitoring the ion fluxes at the target or substrate level \cite{Hecimovic2009,Bohlmark2006,Greczynski2012,Schmidt2013}. The temporal resolution in these studies is often only 20~$\mu$s, which cannot resolve typical pulse lengths between 25~$\mu$s to 50~$\mu$s. Recently, Palmucci et al. succeeded in measurements with a temporal resolution of 2~$\mu$s, but for very short pulses only \cite{Palmucci2013}. These experiments showed that the high kinetic energy of the metal species at the substrate in HiPIMS is responsible for the excellent properties of the deposited films. The energy of these metal species depends at first on the energy due to the sputter process following a Thompson energy distribution. When these fast atoms are ionized in the plasma, the electrical potential at the location of this ionization determines the maximum energy they may possess when reaching the substrate. These electrical potentials have been measured by emissive probe measurements for HiPIMS plasmas and revealed a potential drop of 20~V towards the target in a so called ionization zone (IZ) as proposed by Raadu et al. \cite{Raadu2011}. Therefore, any neutral atoms that are ionized in the IZ experience an electric field pointing towards the target. This leads to the return of the sputtered species and is regarded as an explanation for the poor growth efficiency in HiPIMS plasmas. Another effect is the formation of multiple charged ions \cite{Andersson2008}, which corresponds to a loss channel because energy is dissipated without enhancing the sputter rate. Besides this detrimental effect of multiple charged ions for the growth rate, these species might be beneficial for HiPIMS performance, because they can sustain the high discharge currents by potential electron emission \cite{Andersson2008} in addition to the ohmic heating of the plasma electrons \cite{Huo2013}. Finally, any collisions during the transport from ionization to substrate leads to a thermalization of the energetic species. \\

Several mechanisms have been proposed to explain the presence of energetic metal species in HiPIMS: 

\begin{enumerate}[i)]
\item reflected ions at the target \cite{Palmucci2013}; 
\item the high energy tail of the Thompson distribution \cite{Hecimovic2009}; 
\item negative ions that are generated in front of the target surface that are accelerated towards the substrate by the full target potential \cite{Bowes2013,Sarakinos2010a,Mraz2006a}; 
\item acceleration of metal species by a two-stream instability induced by localized ionization zones on the target \cite{Lundin2008}. Fast optical diagnostics revealed a pattern formation in HiPIMS plasmas, namely the formation of localized ionization zones or so called spokes \cite{Ehiasarian2012,Anders2012}. In the spokes regime the homogeneous plasma torus known from dcMS plasmas changes to a finite number of plasmoids which are rotating over the racetrack. The number of plasmoids is referred to as quasi mode number. These highly localized structures, with a quasi mode number typically between one and four, rotate in the E$\times$B direction in front of the target surface. The velocity of these spokes is ten times smaller than the E$\times$B drift according to single particle motion. The generation mechanism of the spokes and their influence on the transport properties are not fully understood yet. Based on these spokes, energetic ions that have been observed in lateral directions to the magnetron target by Lundin et al. \cite{Lundin2008} who explained these energetic species by a generation mechanism based on a two stream instability interacting with the spoke region;
\item Anders and Andersson et al. \cite{Anders2013,Andersson2013} proposed a model which incorporates a double layer (DL) when the HiPIMS plasma is in the spokes regime. With a DL present the deconfinement and the acceleration of particles to high energies can be explained.
\end{enumerate}

Recently, we investigated the IEDFs in {\it low power} HiPIMS plasmas \cite{Breilmann2013} revealing IEDFs which are dominated by low energy Ti$^+$ ions of a few eV being created from titanium neutrals in front of the substrate. Ti$^+$ ions from the magnetic trap can only reach the substrate after the end of the plasma pulse due to the de-confinement when the plasma is off. All IEDFs could be explained by a Thompson distribution of the sputtered neutrals, which are ionized and thermalized and only slightly shifted in energy due to the acceleration in the sheath in front of the substrate. 

The origin of the very energetic ions in {\it high power} HiPIMS still remains a puzzle. IEDFs with sufficient temporal resolution are necessary to favor one of the presented explanations or models in list i) - v). From time averaged data it is not possbile to determine how the IEDFs evolve or the time when high energy ions are created during the pulse. This constitutes the central goal of this paper. The origin of the energetic metal ions is investigated by combining two fast diagnostics: phase resolved optical emission spectroscopy (PROES) with a temporal resolution of 1~$\mu$s and time- and energy-resolved ion mass spectrometry (MS) with a temporal resolution of 2~$\mu$s. By performing a power scan from dcMS-like to HiPIMS plasmas the different possible mechanisms in producing energetic ions at the substrate are elucidated. We demonstrate that our findings corroborate the proposed potential structure by Anders et al. \cite{Anders2013} and provide evidence how the IEDFs are affected during the pulse.

\section{Experiment}
\subsection{Plasma operation}
A 2-inch magnetron with a titanium target is used as sputter source. The magnetron is powered by a Melec power supply in average-power limited mode to allow stable operation in the runaway regime. The base pressure is $1.5\times10^{-4}$~Pa. Argon 5.0 is used as plasma forming gas at a flow rate of 20~sccm and constant pumping speed. The resulting pressure in the chamber is 0.5~Pa. The pulse length is 50~$\mu$s at a repetition frequency of 300~Hz. The duty cycle is 1.5$\%$ which allows a safe operation of the water-cooled magnetron in the investigated power range.

\begin{figure}[ht]
	\begin{center}
	\includegraphics[width=10cm]{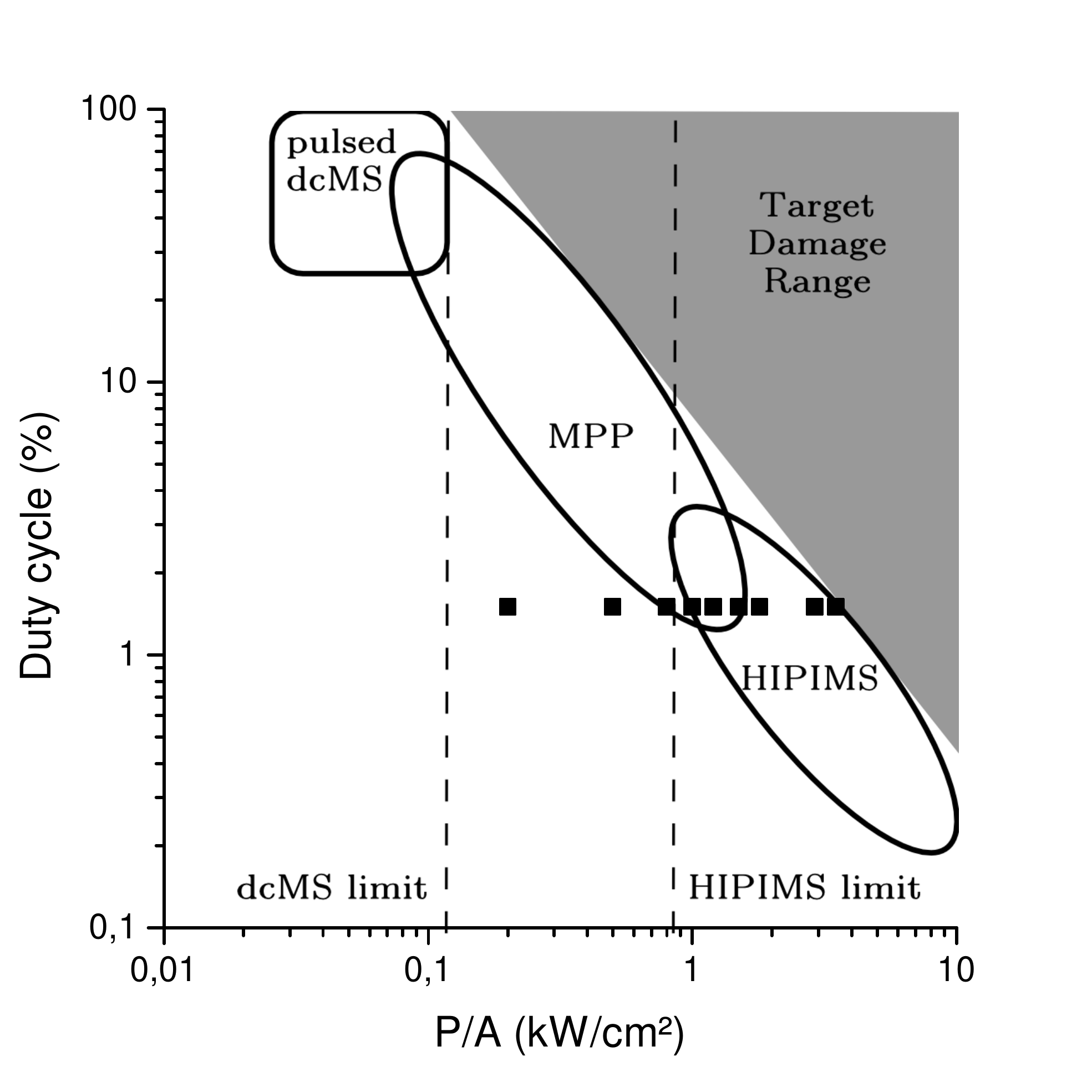}
	\caption{Overview over the conducted experiments and the reachable regimes (dcMS ... direct current magnetron sputtering, MPP ... modulated pulse power ,HiPIMS ...  high power pulsed magnetron sputtering). Background image reproduced from \cite{Gudmundsson2012}.}
	\label{fig:dc_vs_p}
	\end{center}
\end{figure}

The  operating parameters of the plasmas (Fig.~\ref{fig:dc_vs_p}) were selected in a range to cover absorbed power levels from pulsed direct current magnetron sputtering (dcMS), Modulated Pulse Power (MPP) discharges up to HiPIMS \cite{Gudmundsson2012}.

Prior the each measurement, the magnetron was operated at an average power of 0.49~kW for thirty minutes to ensure a thermalization of the experiment as well as a stable chemical composition of the target surface. The experiments were conducted in a single campaign starting with the highest power density of 3.5~kW/cm$^2$ down to the lowest power density of 0.2~kW/cm$^2$. This consecutive campaign reduces the influence of any long term thermal drifts of the experiment as well as of the diagnostic equipment. 

\subsection{Plasma diagnostics}
Voltages and currents are measured directly at the output of the Melec power supply.
The employed VI-probe consists of a LEM LA305-S current transducer with a response time 
$t_r<1$~$\mu$s and a d$I/$d$t>$100~A/$\mu$s and a LEM CV3-1500 voltage transducer with $t_r=0.4$~$\mu$s and d$V/$d$t=$900~V/$\mu$s. The signal damping of the voltage transducer at 800~kHz  is -1~dB at 333~V. All signals are monitored with an oscilloscope (sample interval $dt=$40~ns) and averaged over 128 HiPIMS pulses.\\

Time-resolved ion energy distribution functions (IEDF)  of Ar$^+$, Ar$^{2+}$ and of Ti$^+$, Ti$^{2+}$ are measured using a HIDEN EQP 300 HE instrument. The 100 $\mu$m orifice of the EQP is mounted in line-of-sight facing the racetrack in a distance of 8~cm to the target. The temporal resolution of the diagnostic is 2~$\mu$s. The Low Energetic (LE) part of the IEDFs is measured with an energy resolution of 0.2~eV. The High Energetic (HE) part is monitored with a resolution of 0.5~eV. Data acquisition starts $t$=-15~$\mu$s before the HiPIMS pulse is triggered. The Time of Flight (TOF) of the ions in the EQP has to be accounted for to connect the IEDFs to the measured emission data and VI-probe signals. The TOFs depend on the atomic mass, the charge and the kinetic energy of the ions and were calculated according to the manufacturer specifications, as summarized in Tab.~\ref{tab:eqptof}. All measured signals are proportional to the mass/charge ratio and the energy/charge ratio, respectively. A more detailed description of the setup is given elsewhere \cite{Breilmann2013}.\\

\begin{table}[ht]
	\centering
	\begin{tabular}{|l|c|r|r|r|r|}
	\hline
$P/A$ 		& Ion 			& $E_{min}$	& $E_{max}$	& $TOF_{max}$	& $TOF_{min}$\\
kW/cm$^{2}$	&			& eV		& eV		& $\mu$s	& $\mu$s\\
\hline
0.2	     	& $^{50}$Ti$^{1+}$	& 0		& 10		& 123.3		& 123.1\\
0.2		& $^{48}$Ti$^{2+}$	& 0		& 20		& 85.4		& 85.1\\
0.2		& $^{36}$Ar$^{1+}$	& 0		& 10		& 104.6		& 104.4\\
0.2		& $^{40}$Ar$^{2+}$	& 0		& 10		& 78.0		& 77.8\\
3.5		& $^{50}$Ti$^{1+}$	& 0		& 40		& 123.3		& 122.6\\
3.5		& $^{48}$Ti$^{2+}$	& 0		& 70		& 85.4		& 84.8\\
3.5		& $^{36}$Ar$^{1+}$	& 0		& 10		& 104.6		& 104.4\\
3.5		& $^{40}$Ar$^{2+}$	& 0		& 10		& 78.0		& 77.8\\
\hline
	\end{tabular}
	\caption{Transit times for different ions and isotopes in the HIDEN EQP 300 HE instrument according their charge state and kinetic energy.}
	\label{tab:eqptof}
\end{table}

Simultaneously with the EQP measurements, the optical emission of the plasma is monitored by an intensified CCD camera (ICCD). Bandpass interference filters are used to isolate specific spectral lines. Phase Resolved Optical Emission Spectroscopy (PROES) is performed for Ar neutrals at 760~nm (FWHM 10~nm), for Ar$^+$ ions at 488~nm (FWHM 3~nm), for Ti neutrals at 396~nm (FWHM 3~nm), and for Ti$^+$ ions at 307~nm (FWHM 10~nm). The full width half maxima (FWHM) of the used filters are given in parenthesis. The Field Of View (FOV) is perpendicular to the magnetron axis and covers the region between the top of the anode cover of the magnetron and the lower part of the planar 5 inch substrate holder housing the EQP. The gate time of the camera and the time shift between the images is 1~$\mu$s. The measurement time for every frame is 1~s which corresponds to an averaging over 300 HiPIMS pulses for a particular time step. For further analysis, the intensity profiles on the magnetron axis are extracted for every time frame and presented in a spatio-temporal contour diagram. The intensities for the different species are not calibrated and cannot be compared directly. Any calibration would require an assessment of  the optical paths and of the wavelength dependent sensitivity of the ICCD camera.\\

\section{Results}
\subsection{Voltage Current Measurements}

\begin{figure}[ht]
	\centering
	\includegraphics[width=8cm]{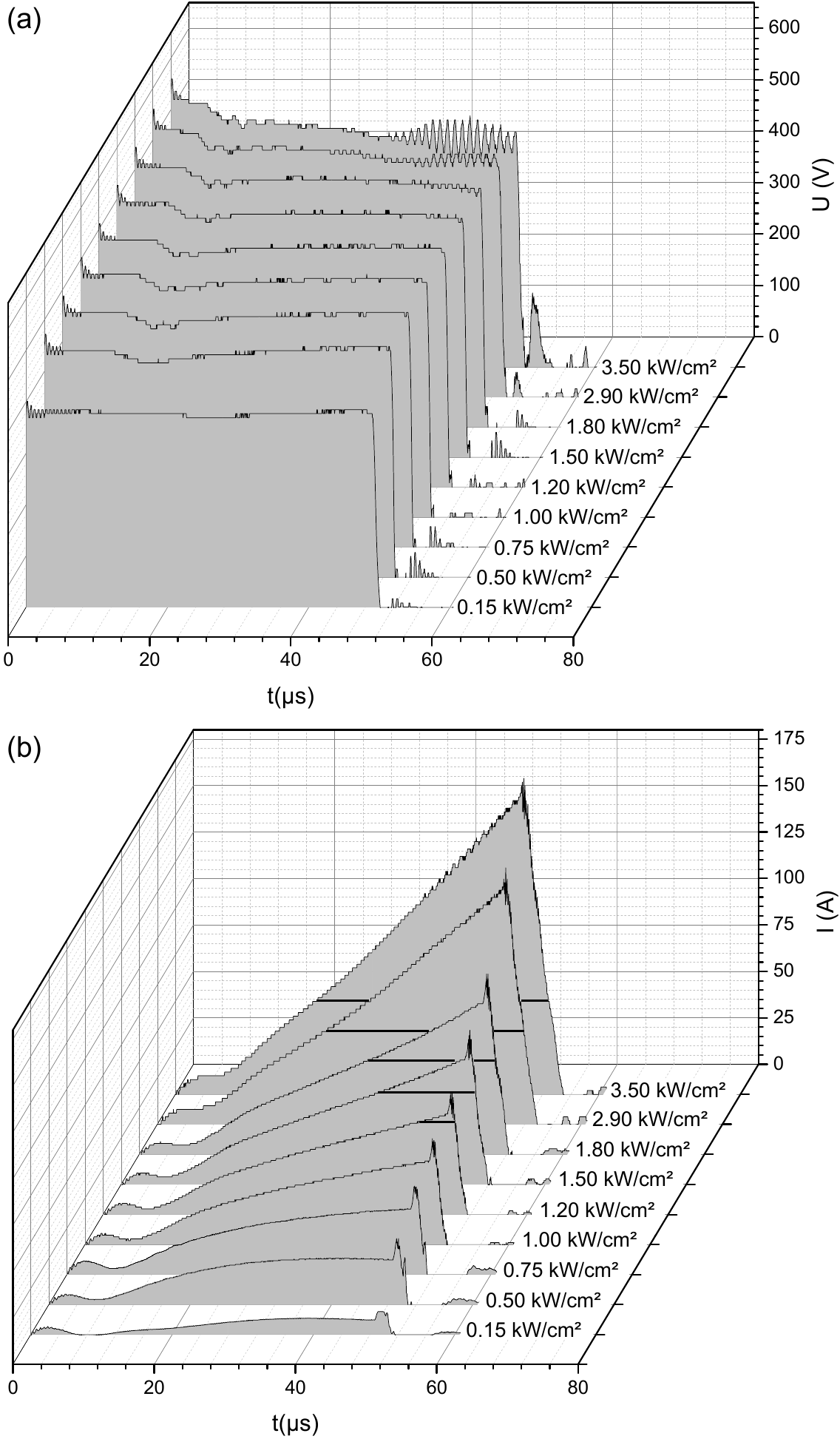}
	\caption{Voltage (a) and current (b) as measured via a VI probe at the power supply. The horizontal solid lines in (b) indicate a current level of 50 A.}
	\label{fig:ivcurves}
\end{figure}

All experiments are monitored by a VI probe attached to the output of the power supply. The resulting VI characteristics are shown in Fig.~\ref{fig:ivcurves}. One can clearly see that the 50~$\mu$s voltage pulses (Fig.~\ref{fig:ivcurves}a) are very rectangular with a maximum voltage decreasing from  -380 V to -500 V with increasing power density. Initially, the voltage at the target adjusts to the set level of the power supply. When the current starts to rise, the voltage  drops by $\approx30$~V and the internal control circuit needs a few $\mu$s to restore the voltage again. For pulses with peak power densities over 1.5~kW/cm$^{2}$ the power supply cannot provide enough current and the voltage decrease only slightly by 10\% (see Fig.~\ref{fig:ivcurves}a). 
The current increases after a typical delay of 10~$\mu$s (Fig.~\ref{fig:ivcurves}b) to a steady state plateau at low power levels. At high power levels it increases continuously until the pulse is switched off. This continuous increase of the current is typical for a HiPIMS plasma entering the self sputtering runaway regime. At power levels above 1~kW/cm$^2$, the HiPIMS regime is reached (Fig.~\ref{fig:dc_vs_p}) and the metallicity of the plasma increases usually because of rarefaction of the noble gas.

It is interesting to note, however, that for the very high power levels, some oscillations in the voltage and the current signals appear at later stages of the HiPIMS pulses. These oscillations differ in frequency and amplitude from the ringing at the beginning of the pulse, which is only a step response of the measurement circuit triggered by the fast rising flanks. Although the VI measurements are averaged over 128 HiPIMS pulses the oscillations at the end of the pulses remain visible and appear to be a robust feature. Those oscillations appear only, if the target current is above $\sim$ 50~A, which is indicated as straight lines in Fig.~\ref{fig:ivcurves}b. This is consistent with direct observations in the literature, where spokes are only observed above a certain current threshold for Ti targets \cite{Winter2013} and Cr, Cu, Nb, Mo Ta targets \cite{Teresa2013}. 

The oscillations exhibit a frequency of 1~MHz, which can be resolved by our VI-probe due to its bandwidth up to 2.5~MHz. Using empirical mode decomposition (EMD) \cite{huang1998}, a signature of these oscillations can also be found in non-averaged signals in the first and second intrinsic mode function (IMF) for the IV-measurements between 1.2~kW/cm$^{2}$ and 3.5~kW/cm$^{2}$. EMD is a technique for analysing non-linear and non-stationary data. With the IMFs and the Hilbert transform, a measured signal $X(t)$ can be expressed as

\begin{equation}
X(t)=\sum_{j=1}^n a_j(t)\exp \Big ( i\int \textrm{d} t\,\omega_j(t)\Big).
\end{equation}

In Fourier decomposition the amplitudes $a_j$ and frequencies $\omega_j$ are constant for every Fourier component $j$. In EMD these quantities are time-dependent. The original signal is therefore decomposed in amplitude modulations $a_j(t)$ and frequency modulations by introducing an instantenous frequency $\omega_j(t)$. Amplitude and frequency modulations are therefore clearly separated and an excellent time resolution can be reached.\\

The occurrence of these 1~MHz oscillations coincides with the appearance of spokes above the HiPIMS target. A characteristic current threshold for their generation has been identified by several authors \cite{Anders2012,Anders2012b,Winter2013,Teresa2013}. These spokes are bright ionization zones which rotate along the racetrack with a typical velocity of 10~kms$^{-1}$. This corresponds to a period of (100~kHz)$^{-1}$ on a 2-inch target. The observed oscillations, however, are  in the range of MHz which corresponds to a period (1~MHz)$^{-1}$. Such a period correlates with the $E\times B$ drift movement of the electrons along the racetrack, which is 10 times faster than the azimuthal velocity of the spoke.\\

To explain the 1~MHz oscillations in the current and voltage signals, we assume the following: the DC Hall current of the electrons rotates along the racetrack with a periodicity of (1~MHz)$^{-1}$. If this Hall current passes the localized ionization zone of a spoke, it is amplified due to the enhanced ionization rate inside the spoke. This local current amplification occurs with a periodicity of the Hall current, which is apparently picked up by the VI probe circuit. Based on this reasoning, we may use the appearance of the 1~MHz oscillations as a signature to identify the formation of spokes in our HiPIMS experiment. Such fast oscillations in the VI signals may be overlooked by other experiments in the literature, because most VI probe electronics in HiPIMS power supplies are not designed for high frequencies. Also, the use of long cables etc. may induce a low pass filter characteristic in the detection chain of the electrical signals which makes those oscillations invisible to VI probe circuits.  

\subsection{Ion energy distributions}

In the following section we present and compare the Ar$^+$, Ar$^{2+}$ and Ti$^+$, Ti$^{2+}$ IEDFs for a peak power density of 3.5~kW/cm$^{2}$ and of 0.2~kW/cm$^{2}$. This corresponds to a comparison between a HiPIMS plasma and a dcMS plasma. The IEDFs are correlated with PROES and VI probe measurements. The roman letters in Figs.~\ref{fig:lear} to \ref{fig:heti} indicate specific features in the data as explained in the following.

\subsubsection{IEDFs at 0.2~kW/cm$^2$}~\\

\begin{figure}[ht]
	\centering
	\includegraphics[width=8cm]{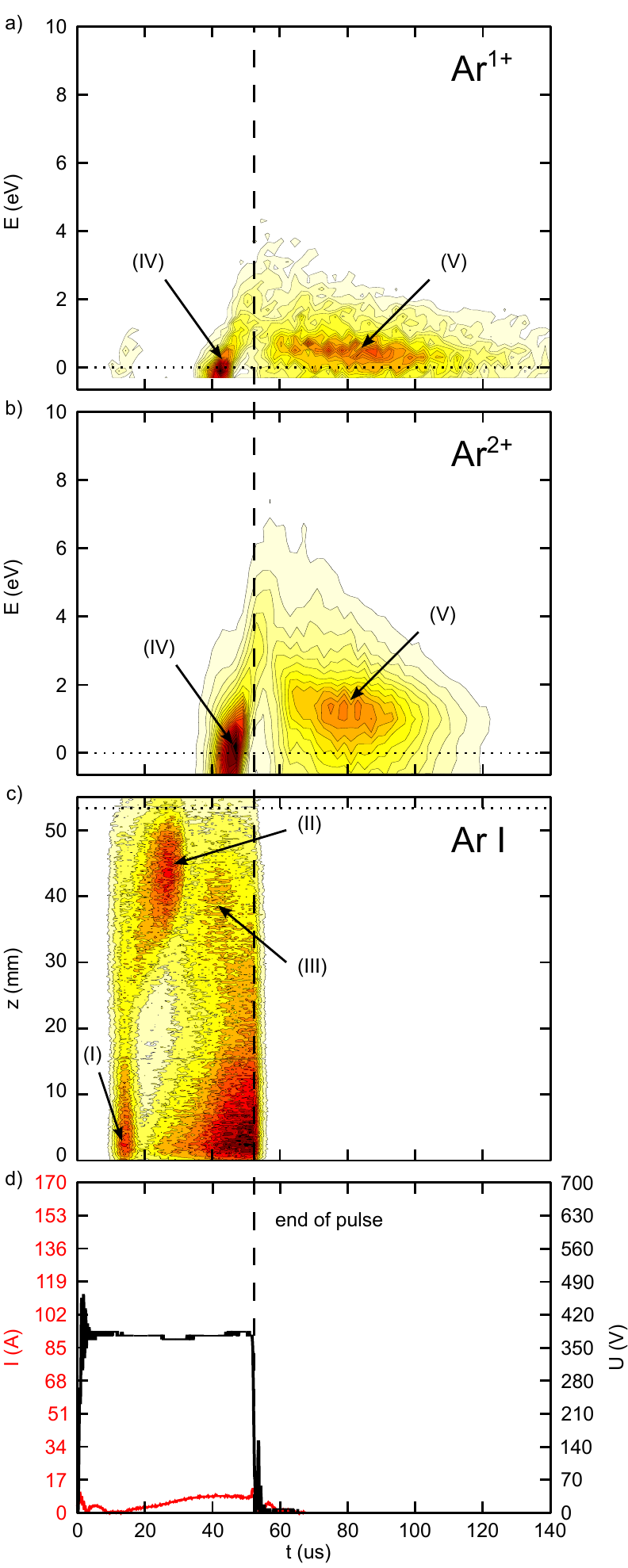}
	\caption{Contour plot of the time dependent ion energy distributions of Ar$^+$ (a) and Ar$^{2+}$ (b). Phase resolved optical emission spectrum (c). Voltage and current of the HiPIMS pulse at a power of 0.2~kW/cm$^2$ (d). The roman letters indicate specific features of the data as explained in the text.}
	\label{fig:lear}
\end{figure}

The time dependence of the  Ar$^+$ and Ar$^{2+}$ IEDFs at a power density of  0.2~kW/cm$^2$ are presented in Fig.~\ref{fig:lear}a and Fig.~\ref{fig:lear}b, respectively. The PROES data in Fig.~\ref{fig:lear}c indicate that the plasma ignites at around 18~$\mu$s. A beam of a hot electrons emerges form the target and causes a distinct excitation of Ar I emission (I). At the same time, an ion acoustic solitary wave \cite{gylfason2005,washimi1966} starts at the target and travels with a velocity of $\approx 3$~km/s towards the substrate (II), where it is reflected \cite{alami2005} inducing a small emission feature (III). 
The energy of Ar$^{+}$ (Fig.~\ref{fig:lear}a) and of Ar$^{2+}$ ions (Fig.~\ref{fig:lear}b) are close to zero at the end of the pulse (IV). This can only be explained, if the plasma in the vicinity of the substrate is cold implying a very small sheath voltage and thus ion energy. The ions in the afterglow are slightly more energetic (V). These are ions being de-trapped from the racetrack\cite{Breilmann2013}. 

The local potential there must be slightly higher, so that the ions have still an energy of 1~eV when they impact on the substrate, although the plasma is off in the afterglow. The energy of Ar$^{2+}$ is twice that of Ar$^{1+}$ which indicates that both species are accelerated in the same electric fields.

\begin{figure}[ht]
	\centering
	\includegraphics[width=8cm]{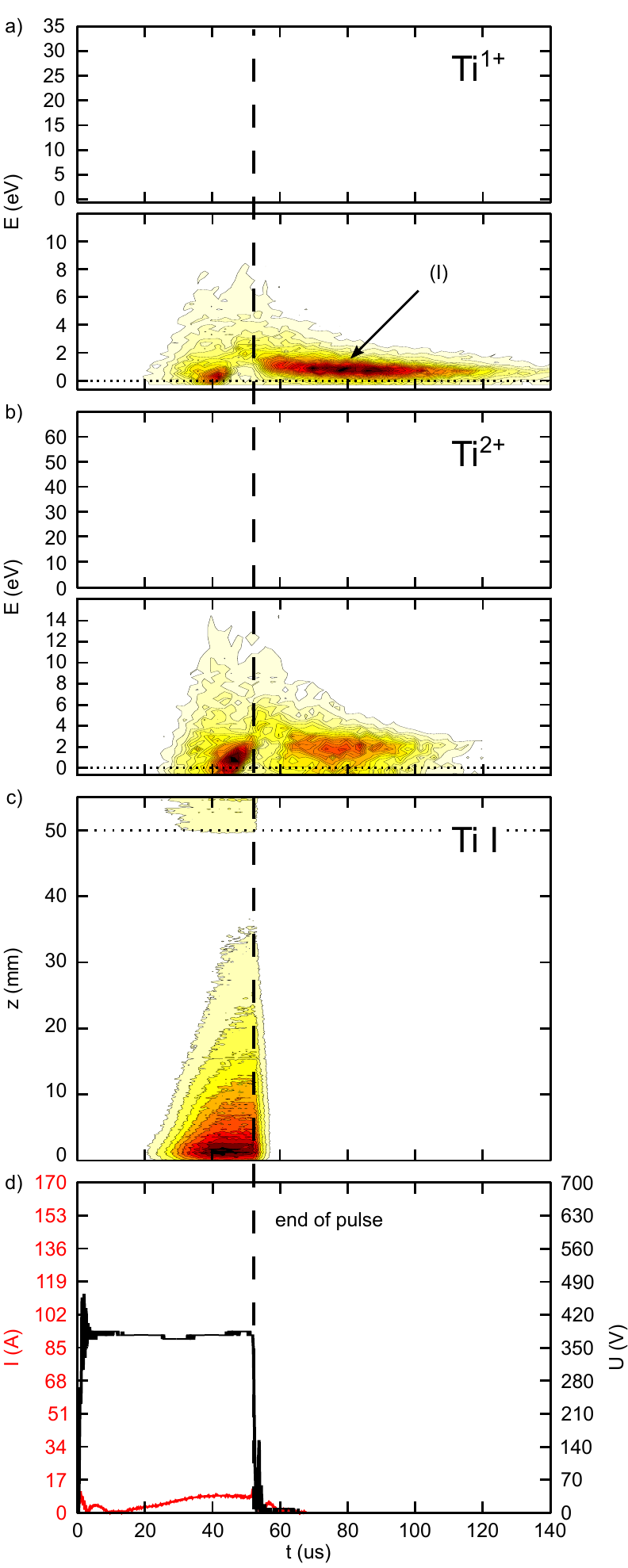}
	\caption{Contour plot of the time dependent ion energy distributions of Ti$^+$ (a) and Ti$^{2+}$ (b). Phase resolved optical emission spectrum (c). The emission at z $>$ 50 originates from a reflection at the substrate holder. Voltage and current of the HiPIMS pulse at a power of 0.2~kW/cm$^2$ (d). The roman letters indicate specific features of the data as explained in the text.}
	\label{fig:leti}
\end{figure}

The time dependence of the  Ti$^+$ and Ti$^{2+}$ IEDFs at a power density of  0.2~kW/cm$^2$ are presented in Fig.~\ref{fig:leti}a and Fig.~\ref{fig:leti}b, respectively. The characteristic of the titanium IEDFs is very similar to that of argon ions. Titanium signals can only be observed later in the pulse in the IEDFs as well as in the PROES measurements. An obvious difference to Fig.~\ref{fig:lear}a,b is that the titanium flux in the afterglow is more pronounced than during the HiPIMS pulse (I). It is also assumed that these ions originate from the confined plasma region where the metalicity of the plasma is high at the end of the pulse. High energy Ti ions are missing in the IEDFs, which corroborates the assumption that the plasma is operated in a pulsed dcMS mode. 

\subsubsection{IEDFs at 3.5 kW/cm$^2$}~\\

\begin{figure}[ht]
	\centering
	\includegraphics[width=8cm]{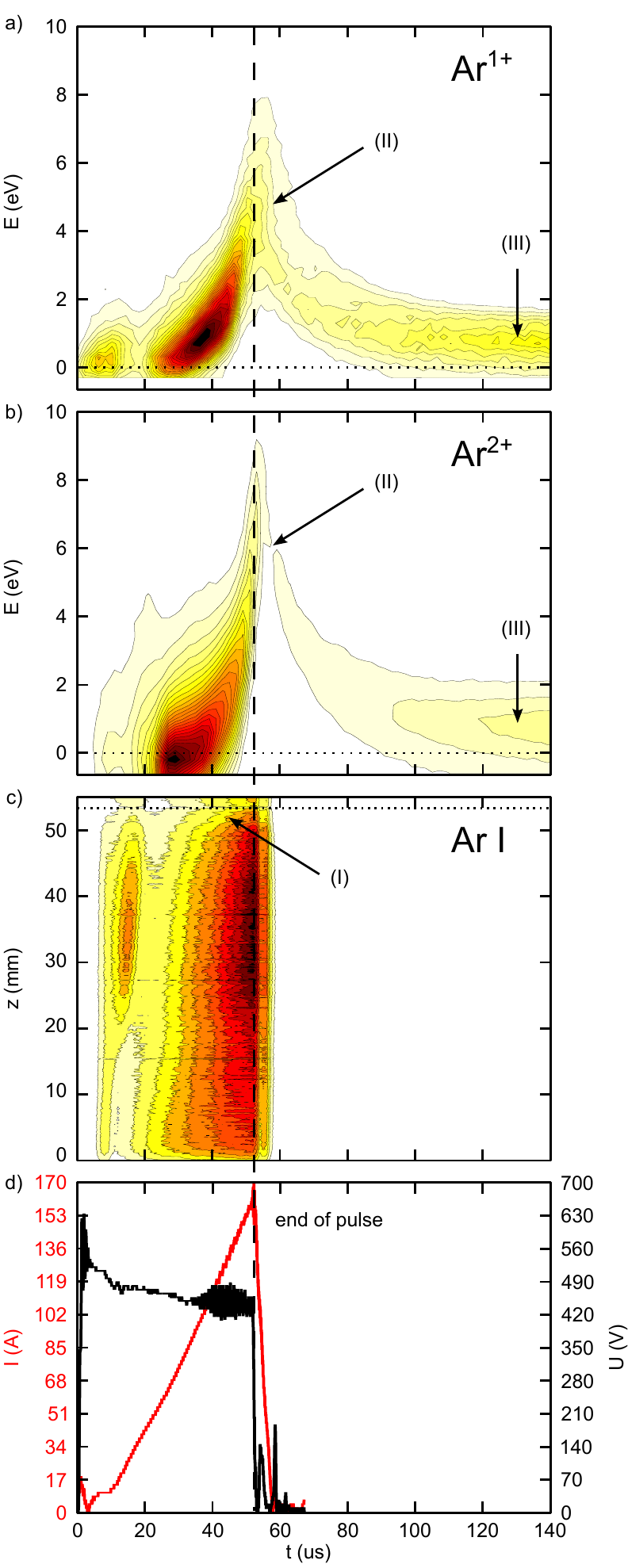}
	\caption{Contour plot of the time dependent ion energy distributions of Ar$^+$ (a) and Ar$^{2+}$ (b). Phase resolved optical emission spectrum (c). Voltage and current of the HiPIMS pulse at a power of 3.5~kW/cm$^2$ (d). The roman letters indicate specific features of the data as explained in the text.}
	\label{fig:hear}
\end{figure}

The time dependence of the  Ar$^+$ and Ar$^{2+}$ IEDFs at a power density of  3.5~kW/cm$^2$ are presented in Fig.~\ref{fig:hear}a and Fig.~\ref{fig:hear}b, respectively. The ignition phase in the high power case, as visible in the PROES image Fig.~\ref{fig:hear}c, is similar to the low power case. The signature of the ion acoustic solitary wave is identical to the low power case. It appears weaker because of the strong Ar I signal at later stages in the pulse. During the pulse the argon ion flux increases and reaches a maximum at the end of the pulse. The energy increases with time. The PROES image further indicates that the plasma is in contact with the substrate Fig.~\ref{fig:hear}c (I). Apparently, the large Hall current at high HiPIMS powers causes a transition to an unbalanced configuration, where the plasma can further extend from the target to the substrate.  The hot plasma in front of the substrate increases the sheath voltage, which accelerates the ions to higher energies than in the low power case.

At the end of the pulse, when the target voltage is switched off and the current ramps down, the IEDFs are shifted by a few eV to higher energies (Fig.~\ref{fig:hear}a,b) (II). This is explained by hot electrons leaving the plasma volume leading to a more positive plasma potential. All ions see this identical potential difference and are accelerated towards the chamber walls and the substrate. This effect has be discussed in detail in \cite{Breilmann2013}.

The argon ion peak in the afterglow (III) is much smaller. Since the HiPIMS plasma is almost in its metallic state, the density of trapped argon ions in the racetrack is expected to be small. Consequently, their contribution to the ion flux at the substrate in the afterglow is small.

\begin{figure}[ht]
	\centering
	\includegraphics[width=8cm]{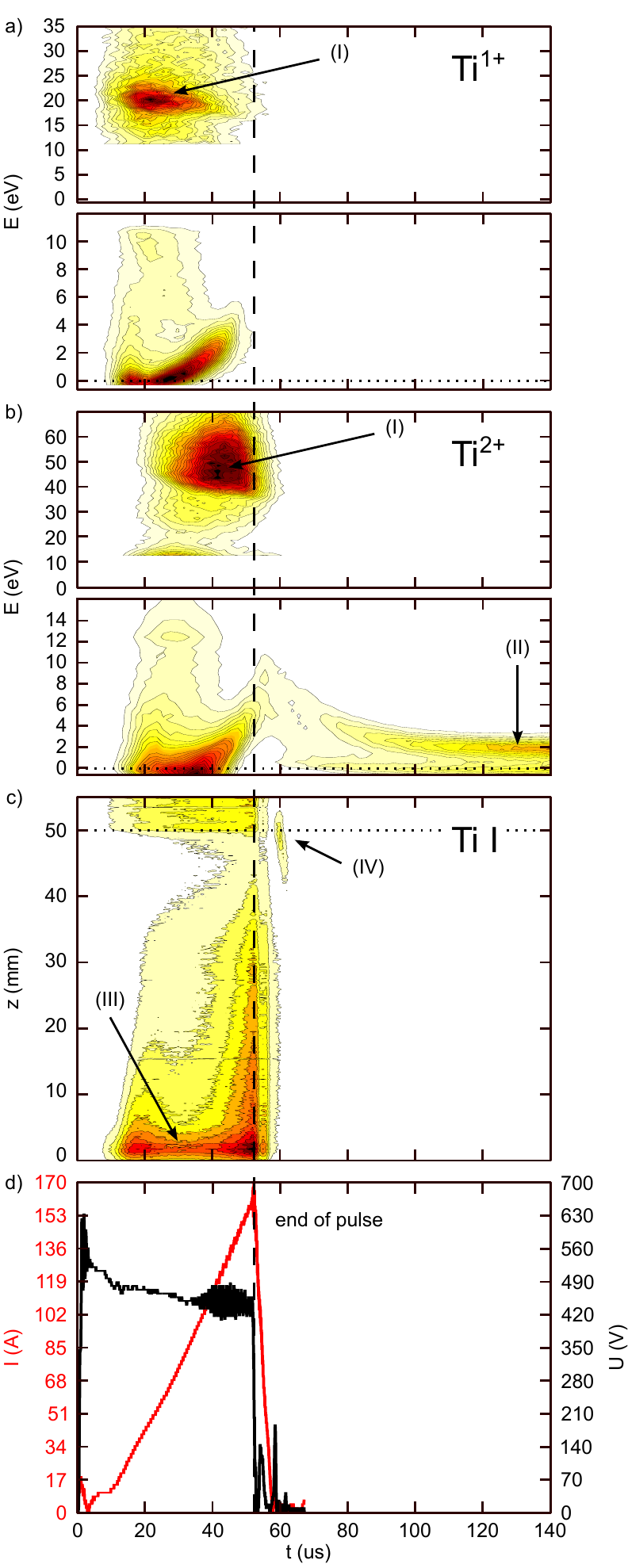}
	\caption{Contour plot of the time dependent ion energy distributions of Ti$^+$ (a) and Ti$^{2+}$ (b). Phase resolved optical emission spectrum (c). The emission at z $>$ 50 originates from a reflection at the substrate holder. Voltage and current of the HiPIMS pulse at a power of 3.5~kW/cm$^2$ (d). The roman letters indicate specific features of the data as explained in the text.}
	\label{fig:heti}
\end{figure}

The time dependence of the  Ti$^+$ and Ti$^{2+}$ IEDFs at a power density of  3.5~kW/cm$^2$ are presented in Fig.~\ref{fig:heti}a and Fig.~\ref{fig:heti}b, respectively. In the Ti ion signals a pronounced high energy (HE) peak shows up (I). This peak appears earlier for Ti$^+$ than for Ti$^{2+}$. This is most likely due to a consecutive ionization sequence from Ti to Ti$^+$ to Ti$^{2+}$ with increasing plasma density and electron temperature. The ion energy of Ti$^{2+}$ is twice that of Ti$^+$ indicating an acceleration of both ions in a region with the same electric field. The characteristics of the Ti LE peaks is similar to the LE peak of argon ions. One can even note that the width of the HE peak of Ti$^{2+}$ broadens with increasing plasma current.

In the afterglow, a characteristic peak of de-trapped ions (II) is only visible for Ti$^{2+}$. This is also consistent, because at the end of the HiPIMS pulse, almost all titanium is converted into Ti$^{2+}$ rendering the Ti$^+$ flux in the afterglow very low. 

The acoustic wave at the beginning of the pulse is missing in the PROES image Fig.~\ref{fig:heti}c. This can easily be explained, because the wave travels through the argon background gas after plasma ignition when no titanium is present yet. In the PROES image one can see, however, that the Ti~I emission increases at first in the beginning, before it decreases, and increases again. This modulation of the Ti I emission is explained by the gas rarefaction window (III): in the  beginning of the HiPIMS pulse, the argon density is large causing an intense sputtering and therefore also a high Ti~I emission signal. When the gas rarefaction sets in after 20~$\mu$s, the argon density is depleted and the lower sputtering rate reduces also the Ti~I emission. When the plasma becomes hotter in front of the target, the Ti~I emission increases again due to the increasing plasma density and electron temperature. In the Ar~I signal in Fig.~\ref{fig:hear}c) this rarefaction window cannot be observed. One has to keep in mind that each  color map is scaled to a maximum intensity which is, in this case, in front of the substrate and might outshine features at low intensity. At low powers the maximum intensity is in front of the target (Fig.~\ref{fig:lear}c). This difference is again attributed to the rarefaction effect. If the onset of sustained self-sputtering is reached, the emission increases almost exponentially due to the avalanche effect of the runaway regime of HiPIMS. After the end of the pulse, a further emission pattern in front of the substrate can be seen (IV), which travels away from the substrate surface. One might speculate that this emission pattern is caused by secondary electrons generated at the substrate by impinging ions at the end of the plasma pulse. These secondary electrons initiate an ion acoustic wave which travels from substrate to target. This wave leads only to a very small emission signature because it propagates in the decaying afterglow plasma.

\subsubsection{Time resolved ion fluxes}~\\

\begin{figure}[ht]
	\centering
	\includegraphics[width=8cm]{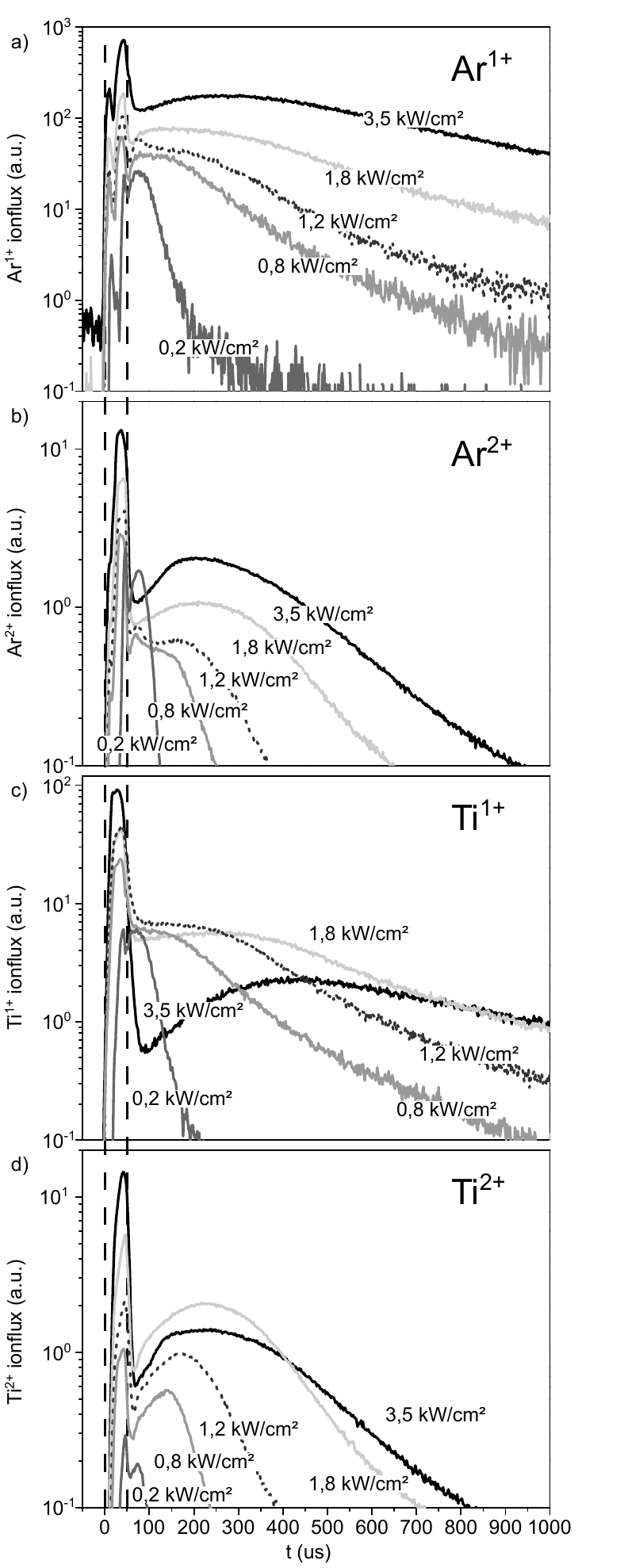}
	\caption{Time dependent ion fluxes of Ar$^+$ (a), Ar$^{2+}$ (b), Ti$^+$ (c) and Ti$^{2+}$ (d) at varying absorbed peak power densities between 0.16~kW/cm$^2$ to 3.5~kW/cm$^2$. The dashed lines indicate the on-time of the plasma pulse.}
	\label{fig:ionfluxes}
\end{figure}

Fig.~\ref{fig:ionfluxes}a,b shows the time dependence of the energy-integrated ion fluxes of Ar$^+$ and Ar$^{2+}$, respectively. The decay times in the afterglow for Ar$^+$ increase with increasing power density. The ion flux at the substrate is a superposition of de-trapped ions from the racetrack region and ions generated in front of the substrate. 
The Ar$^{2+}$ ion flux, however, goes through a pronounced maximum in the afterglow (Fig.~\ref{fig:ionfluxes}b). Ions with higher ionization levels are predominately generated in the intense plasma region above the racetrack where the energy is sufficiently high. This reduces the temporal broadening of the profiles because the fraction of particles generated outside this region is low. Ar$^{2+}$ reaches the substrate much earlier compared to Ar$^+$ after the pulse due to the stronger acceleration in the electric fields. Summarizing, the trapping-detrapping behavior, as discussed in \cite{Breilmann2013} for low power densities, is much more pronounced at high plasma powers. 

Fig.~\ref{fig:ionfluxes}c,d shows the time dependence of the energy-integrated ion fluxes of Ti$^+$ and Ti$^{2+}$, respectively. The trends are similar to that of Ar$^+$ and Ar$^{2+}$. During the pulse on time the maxima of the fluxes grow with the applied power. For Ar$^+$ and Ar$^{2+}$ this is still valid in the afterglow. For titanium at very high power densities (3.5~kW/cm$^{2}$), however, this trend begins to change. The fraction of the incoming ions represent the plasma composition above the racetrack at the end of the pulse. Speculating, high power densities cause a depopulation of Ti$^{1+}$ to even higher ionization levels \cite{Andersson2008}. This effect is less pronounced for Ti$^{2+}$ and not visible for the argon fluxes. Argon has higher ionization energies than titanium and one would expect a depletion of Ar$^+$ at even higher peak power densities.

\section{Discussion}

\subsection{Characteristic of the IEDF}

The measured IEDFs of Ti$^+$, Ti$^{2+}$, Ar$^+$, and Ar$^{2+}$ are summarized as follows. The Ar$^+$ flux exhibits a pronounced peak at the end of the HiPIMS pulse. The Ar$^+$ energy is $\sim1$~eV and increases only slightly with increasing plasma current. In the afterglow, the Ar$^+$ ion flux decreases with a decay time of the order of ms, which is consistent with the very low ion energies close to 0~eV. The Ar$^{2+}$ flux follows mainly the same trends as the Ar$^+$ flux. Only in the afterglow, the Ar$^{2+}$ flux goes through a maximum. Ar$^{2+}$ ions arrive at the substrate earlier than Ar$^+$ ions, which can be explained by the stronger acceleration of Ar$^{2+}$ in the residual ambipolar field of the decaying plasma compared to Ar$^+$. The observation of a pronounced maximum of Ar$^{2+}$ in the afterglow has been explained by the de-confinement of trapped ions after the end of the plasma pulse \cite{Breilmann2013} when the strong potential gradient towards the target vanishes.

\begin{figure}[ht]
	\centering
	\includegraphics[width=12cm]{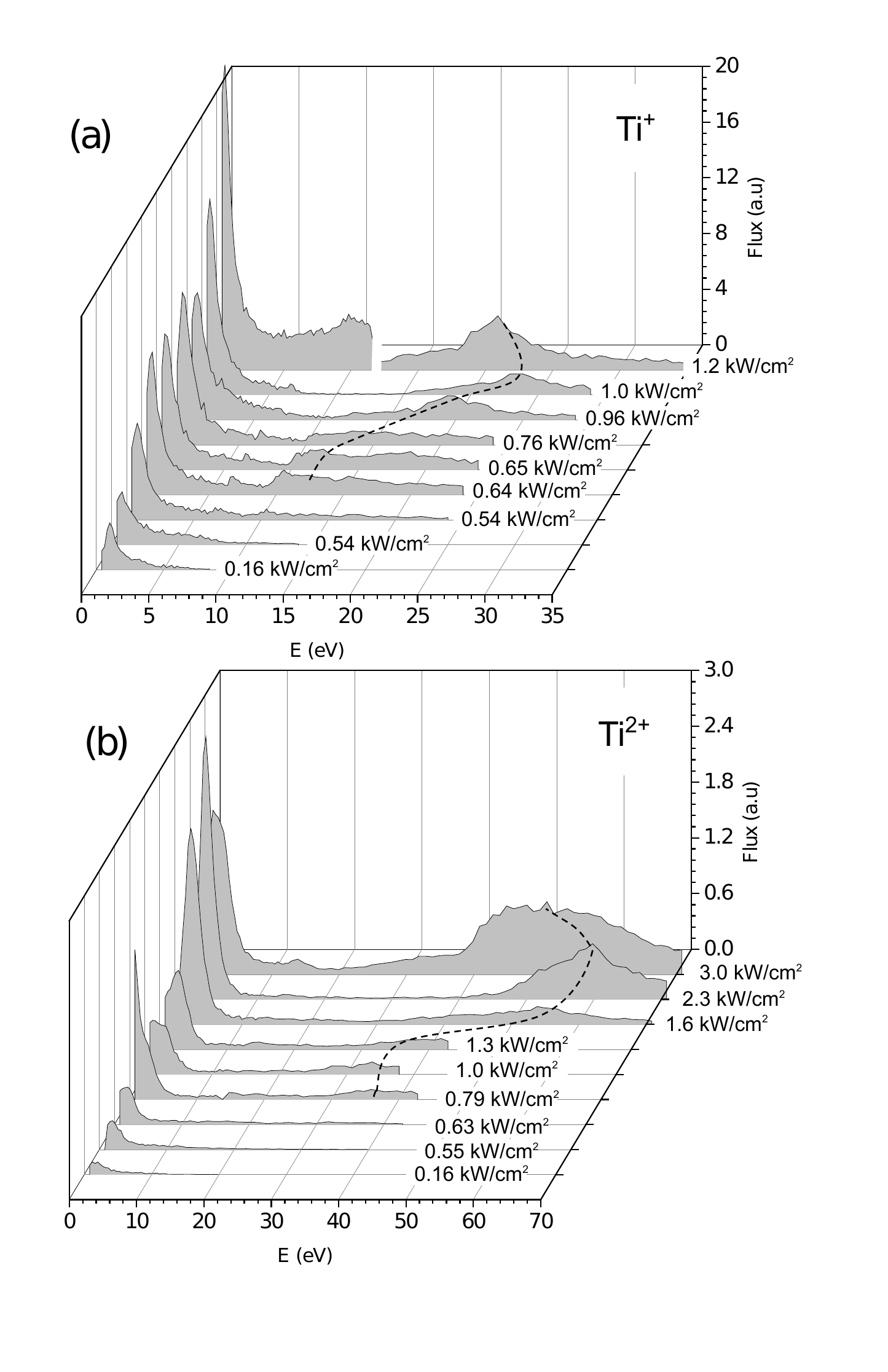}
	\caption{Ion energy distributions of Ti$^+$ (a) and Ti$^{2+}$ (b) at the time of maximum ion flux. The actual plasma power density is indicated. The dashed lines denote the location of the maxima of the {\it high energy} peaks. The discontinuity of Ti$^+$ ion flux at 1.2 ~kW/cm$^{2}$ is due to the fact that two different experiments are combined in this data sets, one distribution for energies 0...10~eV and one measurement from 10~eV$\ldots 35$~eV. Since the detector sensitivity and/or the target wear slowly varies over the time, such discontinuity when combining the spectra of two experiments are expected.}
	\label{fig:iedfti}
\end{figure}

The energy distributions of Ti$^+$ and Ti$^{2+}$ are shown in Fig.~\ref{fig:iedfti} for the moment of maximum ion current during the pulse. One can clearly see, that Ti$^+$ and Ti$^{2+}$ exhibit a distinct peak at low (LE) and high (HE) ion energy. Due to the excellent time resolution in our experiment, we can clearly see that the LE and HE peak can be observed at the {\it same time} during the HiPIMS pulse. As a consequence, any physical model for the explanation of these two distinct peaks needs to provide {\it two distinct} generation mechanisms for the LE and HE peak, which take place simultaneously. This constraint for a physical model was not so stringent before, because most of the IEDF data in the literature were time-integrated or exhibited a rather poor temporal resolution \cite{Bohlmark2006, Lundin2008, Hecimovic2010, Palmucci2013}. The observation of HE peaks in those data could always be explained by a peculiar dynamic of the HiPIMS pulse leading to different energies at different times during the pulse. This is clearly not the case. A physical model to explain the simultaneous observation of LE and HE peaks is developed in the following:  

\begin{itemize}
\item {\it LE peak:} the LE peak corresponds to energies of typically 1~eV, irrespective of the applied plasma power. This is comparable to the energies of Ar$^+$ and Ar$^{2+}$. Titanium species are sputtered at the target and travel towards the substrate with energies according to the Thompson distribution. This distribution has a maximum typically of 3~eV for Ti and a high energy tail. The energy of the Ti$^+$ and Ti$^{2+}$ LE peak at the substrate, however, is much smaller, indicating a thermalization of the species after their ionization in the bulk plasma. The identical LE peaks for argon and for titanium species can easily be explained by a thermalization of both species in the plasma bulk and a residual acceleration of both species in a low voltage sheath in front of the substrate.  

\item{\it HE peak:} at high target power densities, an additional HE peak appears in the titanium IEDFs, with a maximum energy of the order of 15~eV to 25~eV for Ti$^+$ and to the order of 35~eV to 55~eV for Ti$^{2+}$. The energy of the HE peak  increases until a power density of 1.0~kW/cm$^2$ (Ti$^+$) before it decreases again slightly at a power density of 1.2~kW/cm$^2$ (Ti$^{+}$). A similar behaviour is observed for Ti$^{2+}$. Such a HE peak is very distinct from the LE peak and has already been found by several groups \cite{Bohlmark2006, Lundin2008, Hecimovic2010, Palmucci2013}. Its origin is a matter of the current debate and has been explained by several mechanisms:

\begin{itemize}
\item{\it Reflected ions at the target:} Palmucci et al. \cite{Palmucci2013} observed similar distinct peaks in the Ti$^+$ energy distribution for 5~$\mu$s HiPIMS pulses with a maximum current of almost 200~A and a voltage of 1200~V. They speculate that these ions originate from reflected ions at the target. Such a reflection, however, occurs with a probability of typically only 1\%. More important, however, is the fact that such a reflection results in an energy distribution with a maximum at 1~eV, which exponentially decays towards higher energies. This is demonstrated by an SRIM simulation for Ti$^+$ bombarding a Ti surface at an ion energy of 500 eV, as shown in Fig.~\ref{fig:srim}. With this concept, a distinct HE peak for Ti$^+$ and Ti$^{2+}$ cannot be explained.  

\begin{figure}[ht]
	\centering
	\includegraphics[width=8cm]{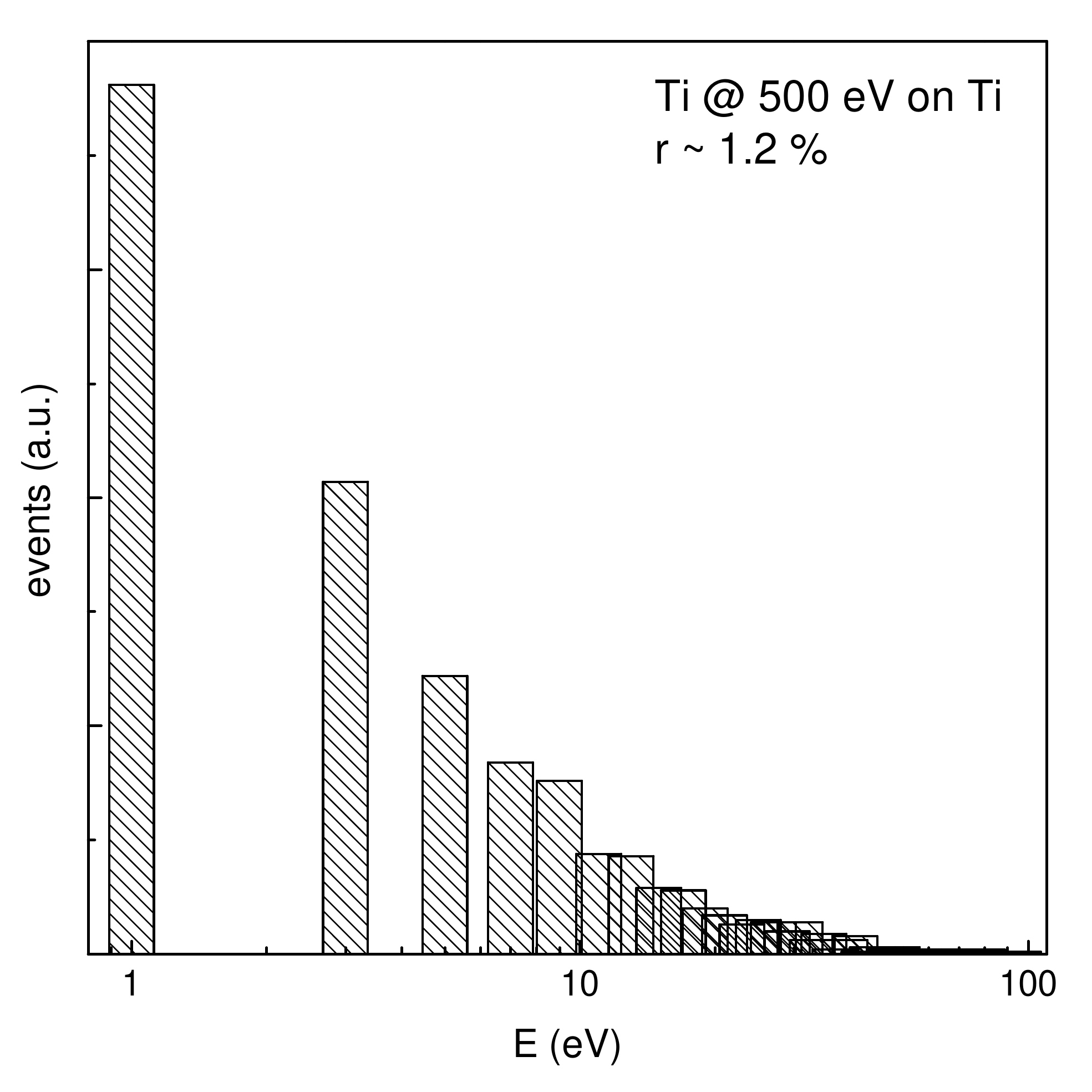}
	\caption{SRIM simulation of Ti$^+$ reflection on Ti for an ion energy of 500~eV. The reflection coefficient is 1.2\%. An energy bin of 1~eV has been chosen for the histogram.}
	\label{fig:srim}
\end{figure}

\item{\it High energy tail of the Thompson distribution:} Hecimovic argued that the HE contribution to the IEDF corresponds to the high energy tail of the Thompson distribution \cite{Hecimovic2010}. The maximum of the energy distribution of sputtered titanium neutrals is only at 3~eV, whereas the HE peak is at 20~eV or even higher. After ionization of the 3~eV neutrals, such ions have to be further accelerated to explain a distinct HE peak of 20~eV. If we assume that this additional acceleration occurs in the sheath in front of the substrate {\it all} ions should exhibit at least the minimum energy corresponding to this sheath voltage. This is not observed. The LE peak and the HE peaks are always observed at the same time. 

\item{\it Signature of the spokes:} Lundin et al. \cite{Lundin2008} observed energetic Ti$^+$ ions in a mass spectrometry measurement monitoring the lateral ion fluxes at the position of the magnetron racetrack. They interpreted this as a signature of plasma spokes and proposed an acceleration mechanism based on a two-stream instability induced by the Hall current above the racetrack. They also observed an azimuthal asymmetry in the IEDFs, which they attributed to the fact that the mass spectrometer monitors the Hall current either clockwise or anti-clockwise, which induces a small energy difference in the IEDFs. Any distinct energy peak in direction of the substrate has been disregarded by Lundin et al. \cite{Lundin2008}, because they argued that the substrate is positioned perpendicular to the direction of the Hall current. 
\end{itemize}

\end{itemize}

This discussion shows that the HE peak in the IEDFs of HiPIMS plasmas remains an unresolved issue. IEDFs in HiPIMS discharges have been reported by many groups revealing their very complex character \cite{Hecimovic2009,Bohlmark2006,Greczynski2012,Schmidt2013}. Depending on the actual plasma process, the IEDF of metal ions is always composed of several contributions: at first, metal atoms are generated by sputtering following a Thompson distribution. This yields energies of a few eV at most. Second, these neutrals are ionized and the plasma potential at this location defines the maximum energy of these ions when they arrive at the substrate. Depending on collision processes between the position of ionization and the substrate, these energetic ions might thermalize, which reduces their energy upon impact at the substrate. The efficiency of this thermalization is not constant in time due to the strong neutral gas dynamic during the HiPIMS pulses. In a typical HiPIMS pulse, the gas rarefaction sets in after typically 20~$\mu$s, which opens a rarefaction windows \cite{Palmucci2013c} in which the neutral gas density is reduced and the transport between target and substrate becomes ballistic. After the end of the pulse, the neutral gas re-fills the depleted plasma volume on a time scale of several 100~$\mu$s. If we take the temporal variation of the plasma potential into account, ion energies might further vary depending on the temporal variation of the electric fields along the ion trajectory from point of ionization until the substrate. As an example, Horwat et al. \cite{Horwat2010b} observed thermalized copper ions with a drift energy from 3~eV to 10~eV depending on the distance to the target. Due to the complex transport properties of the ions from target to substrate, it is difficult to connect the LE and HE peak in the IEDFs to a specific mechanism. This complexity is, however, reduced at very high plasma powers due to gas rarefaction which leads to ballistic transport and reduces the influence of any thermalization. In addition, the electrical fields in high power HiPIMS pulses are strong, which separates individual peaks in the metal IEDFs due to the strong acceleration of the ions. This is exactly the operation window of our experiment, which allows us to isolate the LE and HE peak and their dependence on process parameters. Based on these data we are able to develop a hypothesis on the origin of the HE peaks, as discussed in the following.

\subsection{Hypothesis of double layers as the origin of the HE peaks}

The fact that the Ti$^+$ and Ti$^{2+}$ IEDFs consist of distinct LE and HE peaks being also observed at the same time during the HiPIMS pulse is a strong indication that the ions are generated at very different electrical potentials in the gap between target and substrate. {\it Even more important, the appearance of the HE peaks coincides with the presence of the spokes phenomenon in HIPIMS, if we compare the occurrence of the oscillations in the VI probe measurements in Fig.~\ref{fig:ivcurves} with the Ti IEDFs in Fig.~\ref{fig:heti} }. Apparently, spokes provide a distinct acceleration mechanism of Ti$^+$ and Ti$^{2+}$. Such an acceleration mechanisms may consist of the two-stream instability, as described above, explaining distinct ion energies in lateral directions. But the HE peaks at substrate level are observed in {\it normal direction to the target}. Therefore, we postulate that the Ti$^+$ and Ti$^{2+}$ HE peak originates from a plasma region with {\it positive} plasma potential, which we identify with the high plasma density zone inside a plasma spoke. An identical conclusion has been drawn prior by Anders et al. \cite{Anders2013,Andersson2013} describing it as a ''potential hump'' inside the spoke or as ''propeller blade'' model for HiPIMS to explain the observed lateral ejection of energetic Ti$^+$ ions. Anders et al., however, also concluded that such a potential hump should cause an acceleration of  metal ions in {\it all} directions including the direction towards the substrate. This is exactly observed in our experiment.

The absolute value of the potential hump inside the spoke originates from a double layer (DL) confining the high density plasma region \cite{Charles2009,Raadu1989}. This potential can be estimated by the Boltzmann relation: we assume a typical electron temperature between $T_e=2$~eV and 4~eV and an electron density $n_{sp}$ in the central part of the spoke and a density $n_{bk}$ in the surrounding plasma bulk. We further assume a density ratio $n_{sp}/n_{bk}$ between 10 and 1000. Tab.~\ref{tab:phispoke} summarizes the obtained results for the plasma potential difference $\Phi_{sp}$ according to $n_{sp}/n_{bk}=\exp(\Phi_{sp}/T_e)$. 

\begin{table}[ht]
	\centering
	\begin{tabular}{cr|rrr}
		 			&	& 	\multicolumn{3}{c}{$T_e$~(eV)}  	 \\
					&	& 2 	& 3 	& 4 \\
					\hline
\multirow{3}{*}{\rotatebox{90}{$n_{sp}/n_{bk}$}}	& 10	&  -5	& -7 		& -9 \\
					& 100 	&  -9	& -14   & -18  \\
					& 1000	&  -14	& -21 	&  -28\\
					&	&	&	&		
	\end{tabular}

	\caption{Plasma potential differences $\Phi_{sp}$ for different electron density ratios and electron temperatures according to the Boltzman relation.}
	\label{tab:phispoke}
\end{table}
At higher plasma density ratios and electron temperatures the reachable potential differences are consistent with the measured titanium ion energies. This is only a rough estimate because DLs in a magnetic field are much more complicated since the confinement by the magnetic fields has to be accounted for. To illustrate the formation of the various IEDFs, a sketch of the electrical potential between target and substrate in a HiPIMS discharge is shown in Fig.~\ref{fig:model} for a HiPIMS plasma with (solid line) and without spokes (dashed line), respectively with or without the proposed ''potential hump'':  

\begin{figure}[ht]
	\centering
	\includegraphics[width=8cm]{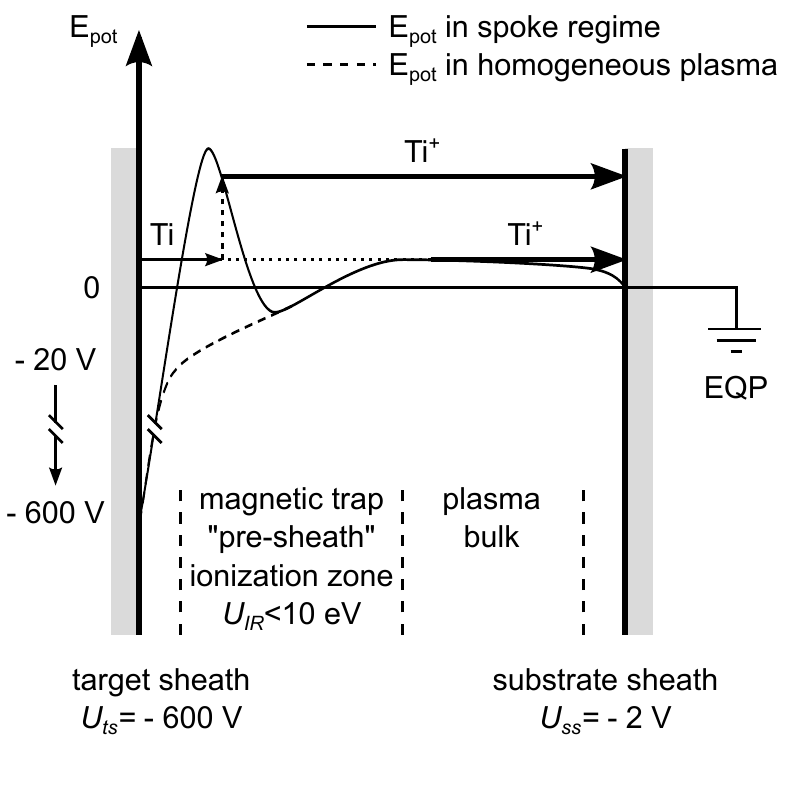}
	\caption{Model for an potential energy of the Ti species between target and substrate in a HiPIMS discharge depending on the electrical potential in the homogenous regime (dashed line) and in the spoke regime (solid line). Titanium species are generated at the target via sputtering at an kinetic energy of typically 3~eV. If they are ionized inside the spoke, they gain the potential energy of the potential hump inside the spoke and may reach the substrate at high energies. If these species are ionized in the plasma bulk, they gain only an energy corresponding to the potential drop in front of the substrate of only a few eV.}
	\label{fig:model}
\end{figure}

\begin{itemize} 

\item {\it HiPIMS plasma with homogeneous plasma torus}: at very low power densities, a homogeneous plasma torus is observed. The sheath voltage in front of the target drops over a very short distance and is of the order of 600~V. Due to the low plasma density in front of the substrate and the low electron temperature, the voltage drop on the substrate side is only 1...2 eV. 
Above the target an ionization zone (IZ) is formed with an extension of a few cm with a voltage drop of typically 20~V towards the target. Such potential distribution has been measured by emissive probes \cite{Rauch2012, Liebig2013} and is consistent with the IZ model of Raadu et al. \cite{Raadu2011}. Titanium neutrals are sputtered at the target surface with a typical energy of 3~eV. If they are ionized inside the IZ, they are accelerated back to the target. This ''return effect'' is believed to be the key obstacle of HiPIMS leading to the reduced growth rate of HiPIMS discharges. Only the high energy tail of the sputtered neutrals according to the Thompson distribution has enough energy to overcome the IZ and to reach the target. Ti neutrals that are ionized inside the plasma bulk reach the substrate and contribute to the LE titanium peak.  

\item {\it HiPIMS plasma in the spokes regime}: at very high power densities, the formation of localized ionization zones or spokes is observed. If we assume a ''potential hump'' inside the spoke region, the energetic trajectory of the sputtered titanium neutrals changes. If the ionization takes place inside the spokes, the Ti$^+$ and Ti$^{2+}$ are accelerated downwards the DL surrounding the spoke, reach the substrate and contribute to the HE peak. If the sputtered titanium species are ionized outside of the spoke region, they either return to the target or they are only accelerated in the low voltage sheath in front of the substrate and contribute to the LE peak. This hypothesis of an acceleration of Ti$^+$ and Ti$^{2+}$ ions in a DL surrounding the spoke is also able to explain the dependence of the IEDFs on the process parameters. 

\begin{itemize}
\item {\it Observation of the HE peak in the spoke regime only}: The correlation of the IV curves Fig.~\ref{fig:ivcurves} with the titanium IEDFs in \ref{fig:heti} showed that the HE peak is only observed at power densities where the HiPIMS plasma is in the spoke mode. The time dependent IEDFs support this correlation. If we regard the Ti$^+$ and Ti$^{2+}$ IEDF in Fig.~\ref{fig:heti}, we see that the HE peak appears only at later stages of the HiPIMS pulse, when the ramp up of the plasma current reaches a value above the threshold necessary for spoke formation.

\item {\it Absence of the HE peak for Ar$^+$ and Ar$^{2+}$}: the spoke phenomenon is a localization of the plasma current in small regions along the racetrack. Due to the local dissipation of the plasma power, the gas depletion is very strong and only the self-sputtering of the metal atoms can provide neutrals to be ionized inside the spoke. Since the neutral background gas argon is expelled from the spoke region, no HE peaks for Ar$^+$ and Ar$^{2+}$ are observed. 

\item {\it Maximum of the HE peak}
Fig.~\ref{fig:iedfti} shows that the energy of the HE peak increases up to 1.0~kW/cm$^{2}$ before it decreases again at 1.2~kW/cm$^{2}$ for Ti$^+$ ions. A similar behaviour is observed for Ti$^{2+}$. The value of the potential hump induced by the DLs depends on the localization of the plasma density inside the spoke. PROES measurements of the spokes revealed that the boundary of the spoke becomes less pronounced at very high powers \cite{Teresa2013}. Consequently, the potential hump is expected to decrease at high powers again. It would be interesting to investigate other materials than titanium, where the plasma becomes homogeneous again at very high power densities, which cannot be reached with our experimental setup.

\item {\it Width of the HE peak}
Fig.~\ref{fig:iedfti} shows that the HE peak becomes broader at higher power densities. The width of the energy distribution depends on {\it spatial} and {\it temporal} broadening mechanisms: (i) the {\it spatial} broadening depends on the location of the ionization event. In a static situation, all ions that are generated inside the spoke experience the full potential drop of the DL. Ions generated inside the DL, experience only a fraction of that acceleration. As a consequence, if the DL has an extension of several cm, a very broad HE peak is expected. The extension of the DL depends on the density differences between plasma bulk and spoke. At higher power densities, this boundary between spoke and plasma bulk becomes less pronounced. As a consequence the difference between $n_{sp}$ and $n_{bk}$ should become smaller and the width of the DL increases. This change of the DL not only reduces the maximum of the HE peak, but also broadens it. (ii) the {\it temporal} broadening depends on the time dependence of the electric fields along the ion trajectory. If we regard an ion traversing the DL it experiences only the full potential drop, if the transit time of the ion through the DL is faster than the DL movement. Since the spoke rotates with a frequency of 100~kHz around the racetrack, this period has to be compared with the transit time. This temporal broadening is identical to the broadening mechanism of IEDFs in RF sheaths \cite{Benedikt2012}, where a frequency of 100~kHz induces a significant broad IEDF. It would be interesting to generate data where the spoke rotation frequency can be freely adjusted and to correlate this to the width of the HE peak. This will be subject of future research. 

\item {\it Energies of Ti$^+$ and Ti$^{2+}$}: due to the acceleration of Ti$^+$ and Ti$^{2+}$ in the same electric fields, the energy of Ti$^{2+}$ is twice that of Ti$^+$. By inspecting the exact energies of the HE peaks of Ti$^+$ and Ti$^{2+}$, one can identify small deviations from this factor two. This can be easily explained, because the location of ionization of Ti$^+$ and Ti$^{2+}$ in space and time are different. Ti$^{2+}$ is usually generated in a stepwise ionization from Ti and Ti$^+$ and is generated therefore at larger distance to the target. At this location, the value of the potential hump might be slightly different leading to small deviations from this factor 2 compared to the HE peak of Ti$^+$. In addition, Ti$^{2+}$ is generated at later times of the HiPIMS pulses, because a higher electron temperature is required for ionization. At this moment in time, the ''potential hump'' might have slightly changed.

\item{\it Asymmetry of the HE peak in lateral directions}: the asymmetry in the lateral direction of the IEDFs, as being reported in the literature\cite{Lundin2008}, can be explained by the Fermi acceleration mechanisms or stochastic heating \cite{Fermi1949,Lieberman1998}, when ions collide with the moving DL surrounding the ''spoke''. If the mass spec monitors the IEDF for a ''spoke'' that moves towards the mass spec, the velocity of the spoke adds to the energy of the HE peak. If the mass spec monitors the IEDF for a ''spoke'' that moves away from the mass spec, the velocity of the spoke reduces the energy of the HE peak. 
\end{itemize}
\end{itemize}

This discussion illustrates that the acceleration of metal ions in DLs around the spokes in HiPIMS plasma is able to explain consistently all experiments in this paper as well as experimental observations in the literature. It is surprising that such a potential hump has not yet been measured by probe diagnostics. One has to keep in mind, however, that probes are invasive and might interfere with the spokes or even suppress their formation. This might be the reason, why the existing plasma potential measurements using emissive probes have collected data always outside the plasma torus \cite{Rauch2012, Liebig2013}. In addition, any probe diagnostic has to sample the dynamic spoke with a sample rate of at least 100~kHz to resolve the potential variations. 

Based on this hypothesis, the spoke phenomenon provides a mechanism to release the confined Ti$^+$ ions in the magnetic trap of the HiPIMS plasma resulting in an energetic Ti ion beam towards the substrate. Without the spoke being present, the generated ions return to the target, or reach the substrate only in the afterglow at very low ion energies.  

\section{Conclusion}

The IEDFs in a HiPIMS process for titanium sputtering consist of a low energy (LE) peak for titanium and argon ions and of an additional high energy (HE) peak for titanium ions only. Those HE titanium ions are only generated when the HiPIMS plasma enters the spoke regime. It is proposed that the acceleration of Ti$^+$ ions leading to the HE peak occurs inside a double layer (DL) surrounding the spoke, namely a high plasma density region. These energetic titanium ions are ejected in all directions from the spoke, which allows the Ti ions to escape the magnetic trap of the confining HiPIMS plasma. This mechanism can consistently explain all dependencies in this experiments as well as observations in the literature.

The IEDF of  metal species arriving at the substrate defines the energy input during film growth, which is of paramount importance for all film properties. Since the DL around the spokes provides a mechanism to overcome the return effect in HiPIMS and to provide energetic Ti$^+$ species, one may conclude that the spoke phenomenon is not a nuisance or peculiarity of the HiPIMS process, but rather the {\it essence} of HiPIMS plasmas explaining their good performance for material synthesis applications.

It will be extremely important in the future to directly measure the plasma potentials and electron densities {\it inside} the plasma spoke with fast non-invasive techniques. Such diagnostic techniques have to be synchronized with the 100~kHz movement of the spokes, which underlines the huge experimental challenges for such measurements. In addition, any further theoretical description of the phenomenon in HiPIMS plasma has to include kinetic effects of species acceleration and the large internal electric fields in the postulated double layers.

\section*{Acknowledgements}

The authors would like to thank Norbert Grabkowski and Felix Mitschker for their
technical support. This project is supported by the DFG (German
Science Foundation) within the framework of the Coordinated
Research Center SFB-TR 87 and the Research Department ''Plasmas
with Complex Interactions'' at Ruhr-University Bochum. 

\section*{References}
\bibliography{bibmac} 

\providecommand{\newblock}{}
\begin{thebibliography}{10}
\expandafter\ifx\csname url\endcsname\relax
  \def\url#1{{\tt #1}}\fi
\expandafter\ifx\csname urlprefix\endcsname\relax\def\urlprefix{URL }\fi
\providecommand{\eprint}[2][]{\url{#2}}

\bibitem{Sarakinos2010}
Sarakinos K, Alami J and Konstantinidis S 2010 {\em Surface and Coatings
  Technology\/} {\bf 204} 1661--1684 ISSN 02578972

\bibitem{Gudmundsson2012}
Gudmundsson J~T, Brenning N, Lundin D and Helmersson U 2012 {\em Journal of
  Vacuum Science {\&} Technology A: Vacuum, Surfaces, and Films\/} {\bf 30}
  030801

\bibitem{Mishra2010}
Mishra A, Kelly P~J and Bradley J~W 2010 {\em Plasma Sources Science and
  Technology\/} {\bf 19} 045014 ISSN 0963-0252

\bibitem{Mishra2011}
Mishra A, Kelly P~J and Bradley J~W 2011 {\em Journal of Physics D: Applied
  Physics\/} {\bf 44} 425201 ISSN 0022-3727

\bibitem{Lundin2011}
Lundin D, Sahab S~A, Brenning N, Huo C and Helmersson U 2011 {\em Plasma
  Sources Science and Technology\/} {\bf 20} 045003 ISSN 0963-0252

\bibitem{Lundin2009}
Lundin D, Brenning N, J{\"a}dern{\"a}s D, Larsson P, Wallin E, Lattemann M,
  Raadu M~A and Helmersson U 2009 {\em Plasma Sources Science and Technology\/}
  {\bf 18} 045008 ISSN 0963-0252

\bibitem{Hala2010}
Hala M, Viau N, Zabeida O, {Klemberg-Sapieha, J E} and Martinu L 2010 {\em
  Journal of Applied Physics\/} {\bf 107} 043305

\bibitem{Anders2011}
Anders A 2011 {\em Surface and Coatings Technology\/} {\bf 205} S1--S9 ISSN
  02578972

\bibitem{Poolcharuansin2010}
Poolcharuansin P and Bradley J~W 2010 {\em Plasma Sources Science and
  Technology\/} {\bf 19} 025010 ISSN 0963-0252

\bibitem{Hecimovic2009}
Hecimovic A and Ehiasarian A~P 2009 {\em Journal of Physics D: Applied
  Physics\/} {\bf 42} 135209

\bibitem{Bohlmark2006}
Bohlmark J, Lattemann M, Gudmundsson J~T, Ehiasarian A~P, Aranda~Gonzalvo Y,
  Brenning N and Helmersson U 2006 {\em Thin Solid Films\/} {\bf 515}
  1522--1526 ISSN 00406090

\bibitem{Greczynski2012}
Greczynski G, Lu J, Jensen J, Petrov I, Greene J~E, Bolz S, K{\"o}lker W,
  Schiffers C, Lemmer O and Hultman L 2012 {\em Journal of Vacuum Science {\&}
  Technology A: Vacuum, Surfaces, and Films\/} {\bf 30} 061504 ISSN 07342101

\bibitem{Schmidt2013}
Schmidt S, Czig{\'a}ny Z, Greczynski G, Jensen J and Hultman L 2013 {\em
  Journal of Vacuum Science {\&} Technology A: Vacuum, Surfaces, and Films\/}
  {\bf 31} 011503 ISSN 07342101

\bibitem{Palmucci2013}
Palmucci M, Britun N, Silva T, Snyders R and Konstantinidis S 2013 {\em Journal
  of Physics D: Applied Physics\/} {\bf 46} 215201 ISSN 0022-3727

\bibitem{Raadu2011}
Raadu M~A, Axn{\"a}s I, Gudmundsson J~T, Huo C and Brenning N 2011 {\em Plasma
  Sources Science and Technology\/} {\bf 20} 065007 ISSN 0963-0252

\bibitem{Andersson2008}
Andersson J, Ehiasarian A~P and Anders A 2008 {\em Applied Physics Letters\/}
  {\bf 93} 071504

\bibitem{Huo2013}
Huo C, Lundin D, Raadu M~A, Anders A, Gudmundsson J~T and Brenning N 2013 {\em
  Plasma Sources Science and Technology\/} {\bf 22} 045005 ISSN 0963-0252

\bibitem{Bowes2013}
Bowes M, Poolcharuansin P and Bradley J~W 2013 {\em Journal of Physics D:
  Applied Physics\/} {\bf 46} 045204 ISSN 0022-3727

\bibitem{Sarakinos2010a}
Sarakinos K, Music D, Mr{\'a}z S, to~Baben M, Jiang K, Nahif F, Braun A,
  Zilkens C, Konstantinidis S, Renaux F, Cossement D, Munnik F and Schneider
  J~M 2010 {\em Journal of Applied Physics\/} {\bf 108} 014904

\bibitem{Mraz2006a}
Mr{\'a}z S and Schneider J~M 2006 {\em Applied Physics Letters\/} {\bf 89}
  051502 ISSN 00036951

\bibitem{Lundin2008}
Lundin D, Larsson P, Wallin E, Lattemann M, Brenning N and Helmersson U 2008
  {\em Plasma Sources Science and Technology\/} {\bf 17} 035021 ISSN 0963-0252

\bibitem{Ehiasarian2012}
Ehiasarian A~P, Hecimovic A, {de los Arcos, T}, New R, {Schulz-von der Gathen,
  V}, B{\"o}ke M and Winter J 2012 {\em Applied Physics Letters\/} {\bf 100}
  114101

\bibitem{Anders2012}
Anders A 2012 {\em Applied Physics Letters\/} {\bf 100} 224104

\bibitem{Anders2013}
Anders A, Panjan M, Franz R, Andersson J and Ni P 2013 {\em Appl. Phys.
  Lett.\/} {\bf 103} 144103

\bibitem{Andersson2013}
Andersson J, Ni P and Anders A 2013 {\em Applied Physics Letters\/} {\bf 103}
  054104 ISSN 00036951

\bibitem{Breilmann2013}
Breilmann W, Maszl C, Benedikt J and Keudell A~v 2013 {\em Journal of Physics
  D: Applied Physics\/} {\bf 46} 485204 ISSN 0022-3727

\bibitem{Winter2013}
Winter J, Hecimovic A, {de los Arcos, T}, B{\"o}ke M and {Schulz-von der
  Gathen, V} 2013 {\em Journal of Physics D: Applied Physics\/} {\bf 46} 084007
  ISSN 0022-3727

\bibitem{Teresa2013}
{Arcos, T de los}, Layes V, Gonzalvo Y~A, {Gathen, V Schulz-von der}, Hecimovic
  A and Winter J 2013 {\em Journal of Physics D: Applied Physics\/} {\bf 46}
  335201 ISSN 0022-3727

\bibitem{huang1998}
Huang N~E, Shen Z, Long S~R, Wu M~C, Shih H~H, Zheng Q, Yen N~C, Tung C~C and
  Liu H~H 1998 {\em Proceedings of the Royal Society A: Mathematical, Physical
  and Engineering Sciences\/} {\bf 454} 903--995 ISSN 1364-5021

\bibitem{Anders2012b}
Anders A, Ni P and Rauch A 2012 {\em Journal of Applied Physics\/} {\bf 111}
  053304

\bibitem{gylfason2005}
Gylfason K~B, Alami J, Helmersson U and Gudmundsson J~T 2005 {\em Journal of
  Physics D: Applied Physics\/} {\bf 38} 3417--3421 ISSN 0022-3727

\bibitem{washimi1966}
Washimi H and Taniuti T 1966 {\em Physical Review Letters\/} {\bf 17} 996--998
  ISSN 0031-9007

\bibitem{alami2005}
Alami J, Gudmundsson J~T, Bohlmark J, Birch J and Helmersson U 2005 {\em Plasma
  Sources Science and Technology\/} {\bf 14} 525--531 ISSN 0963-0252

\bibitem{Hecimovic2010}
Hecimovic A and Ehiasarian A~P 2010 {\em Journal of Applied Physics\/} {\bf
  108} 063301 ISSN 00218979

\bibitem{Palmucci2013c}
Palmucci M, Britun N, Konstantinidis S and Snyders R 2013 {\em Journal of
  Applied Physics\/} {\bf 114} 113302 ISSN 00218979

\bibitem{Horwat2010b}
Horwat D and Anders A 2010 {\em Applied Physics Letters\/} {\bf 97} 221501

\bibitem{Charles2009}
Charles C 2009 {\em Journal of Physics D: Applied Physics\/} {\bf 42} 163001
  ISSN 0022-3727

\bibitem{Raadu1989}
Raadu M~A 1989 {\em Physics Reports\/} {\bf 178} 25--97 ISSN 03701573

\bibitem{Rauch2012}
Rauch A, Mendelsberg R~J, Sanders J~M and Anders A 2012 {\em Journal of Applied
  Physics\/} {\bf 111} 083302

\bibitem{Liebig2013}
Liebig B and Bradley J~W 2013 {\em Plasma Sources Science and Technology\/}
  {\bf 22} 045020 ISSN 0963-0252

\bibitem{Benedikt2012}
Benedikt J, Hecimovic A, Ellerweg D and Keudell A~v 2012 {\em Journal of
  Physics D: Applied Physics\/} {\bf 45} 403001

\bibitem{Fermi1949}
Fermi E 1949 {\em Physical Review\/} {\bf 75} 1169--1174 ISSN 0031-899X

\bibitem{Lieberman1998}
{Lieberman, M a} and Godyak V~A 1998 {\em IEEE Transactions on Plasma
  Science\/} {\bf 26} 955--986

\end{thebibliography}

\end{document}